\documentclass[twocolumn,floatfix,aps,prd,showpacs,amsmath,amssymb,nofootinbib,preprintnumbers,superscriptaddress]{revtex4}

\usepackage{graphicx}

\newcommand{\simg}{\gtrsim}
\newcommand{\siml}{\lesssim}

\begin{document}

\title{Precision calculations of the gravitational wave background spectrum from inflation}

\author{Sachiko Kuroyanagi} \email[]{s-kuro@a.phys.nagoya-u.ac.jp}
\affiliation{Department of Physics, Nagoya University, Chikusa, Nagoya
  464-8602, Japan}

\author{Takeshi Chiba} \affiliation{Department of Physics, College of
  Humanities and Sciences, Nihon University, Tokyo 156-8550, Japan}

\author{Naoshi Sugiyama} \affiliation{Department of Physics, Nagoya
  University, Chikusa, Nagoya 464-8602, Japan} \affiliation{Institute
  for Physics and Mathematics of the Universe, University of Tokyo,
  5-1-5 Kashiwa-no-ha, Kashiwa City, Chiba 277-8582, Japan}

\begin{abstract}

The spectrum of the gravitational wave background originating from
quantum fluctuations during inflation is calculated numerically for
various inflation models over a wide range of frequencies. We take into
account four ingredients : the scalar field dynamics during inflation
making no use of the slow-roll approximation, the fermionic decay of the
scalar field with a small coupling constant during the reheating
process, the change of the effective number of degrees of freedom $g_*$
in the radiation-dominated era, and the anisotropic stress of
free-streaming neutrinos. By numerically solving the evolution of
gravitational waves during and after inflation up to the present, all of
these effects can be examined comprehensively and accurately over a
broad spectrum, even at very high frequencies. We find that the spectrum
shows (i) a large deviation from the spectrum less accurate obtained by
Taylor expanding around the CMB scale using the slow-roll approximation
(ii) a characteristic frequency dependence due to the reheating effect,
and (iii) damping due to the $g_*$ changes and the neutrino anisotropic
stress. We suggest that future analysis of the gravitational wave
background should take into consideration the fact that analytical
estimates using the Taylor expansion overestimate the amplitude of the
spectrum.
\end{abstract}


\maketitle

\section{Introduction}

Inflation \cite{inf1,inf2,inf3} is now widely considered to be the
most attractive scenario for describing the early evolution of the
Universe. One of the reasons why inflation succeeds as a standard
paradigm is that it not only solves conundrums in the big bang theory
(horizon/flatness/monopole problems) but also predicts an almost
scale-invariant spectrum of scalar (density) perturbations originating
from quantum fluctuations \cite{dens1,dens2,dens3}. This theoretical
prediction is consistent with large-scale structure and cosmic
microwave background (CMB) observations, and there have been intensive
attempts to obtain a deeper understanding of inflation dynamics
through analysis of these observations. Moreover, inflation also
generates an almost scale-invariant spectrum of tensor perturbations
(gravitational waves) \cite{grav1,grav2,grav3}, the detection of which
is expected to provide a new source of information on the early
Universe, and provide a key to discriminate between inflation and
alternative models.  There are ongoing efforts to detect indirectly
this inflation-produced gravitational wave background, such as the
next-generation CMB experiments which measure CMB polarization
patterns induced by gravitational waves \cite{CMBp1,CMBp2}. Clearly,
the direct detection \cite{lisa,decigo,bbo,ligo2,Kuroda,Nishizawa} of
the primordial gravitational wave background would provide an even
more attractive prospect.

The most striking property of gravitational waves is their weakness of
interaction with matter. While this property makes direct detection very
difficult, it may provide the following specific benefits. Since
primordial gravitational waves propagate freely since the end of
inflation because of their weak interactions with matter, one may
observe a remnant of inflation directly through observation. In
addition, due to the smallness of their amplitude, a linear
approximation is valid for gravitational waves in all scales.  This
makes it easy to study their behavior at any frequency.  This is in
contrast to density perturbations, whose evolution is more complicated,
and which have now grown nonlinear at small scales.  [Although
gravitational waves are affected by the second order term of the
primordial density perturbations \cite{Baumann} and the growing
nonlinear modes \cite{Carbone} through the anisotropic stress term,
these effects appear at very low frequencies \footnote{One might expect
more gravitational waves are generated on smaller scales due to the
nonlinear structure formation since the structure becomes nonlinear
from the smaller scales in the CDM model.  However, it is not the case
because the amplitude of the gravitational waves generated at high
redshift decreases while they propagate in the expanding Universe, and
becomes negligible small. (See Ref. \cite{Carbone}, for details.)}
($\siml 10^{-15}$Hz) and have been estimated to be small.]  Therefore,
the gravitational wave background has the potential to allow us to
extract "clean" information from the early Universe, and is extremely
attractive as a tool of future observational cosmology.

In this paper, our goal is to provide precise predictions for the
spectrum amplitude of the gravitational wave background as preparation
for assessing detectability in future experiments aiming at the direct
detection of gravitational waves. It is a well-known fact that the tilt
and the amplitude of the spectrum strongly depend on the inflation
model. Also, after inflation, changes in the Hubble expansion rate and
the anisotropic stress of neutrinos induce characteristic features in
the spectrum \cite{Schwarz, Weinberg}. These components which affect the
spectrum should be considered all together to estimate accurately the
amplitude of the spectrum over a wide frequency range. The latter
effects have been well investigated at frequencies below $10^{-2}$Hz by
Watanabe and Komatsu \cite{Watanabe}, who calculate the spectrum taking
into account both the change of the effective number of degrees of
freedom and the neutrino anisotropic stress. In contrast, our interest
lies particularly in the higher-frequency region of the gravitational
wave background spectrum.

The reason why we focus on the high frequency modes is that they are
sensitive to a crucial aspect of the dynamics of inflation: how
inflation terminates. The high frequency modes are produced outside the
horizon near the end of inflation and reenter the horizon soonest after
inflation ends, in the period where reheating is considered to occur.
Since the amplitude of a gravitational-wave mode is affected by the
Hubble expansion rate at the point when that mode crosses the horizon,
the spectrum of these modes depends on the particular inflation and
reheating models chosen. This means that it may be possible to
extract information about the epoch, which is still quite unclear,
through direct detection of the gravitational wave background.

In this paper, we numerically analyze the spectrum of the primordial
gravitational wave background for several inflation models. It is
particularly worth noting that the dynamics of inflation and reheating
are included by numerically solving the relevant equations, which
describe an inflaton field rolling down its potential and subsequently
decaying into radiation with oscillation at the bottom of the
potential. This means that we make no use of the slow-roll
approximation, which is no longer valid near the end of inflation,
unlike many previous works \cite{Boyle,Turner,Smith,Friedman}. Our
numerical approach for computing the dynamics of inflation enables us to
find the deviation from this approximation, which depends strongly on
the inflation model. Moreover this numerical approach works effectively
to connect the change of the Hubble expansion rate smoothly from the
inflation phase to the radiation-dominated era via the reheating
process, which cannot be accurately treated analytically. Since the
Hubble expansion rate depends on the behavior of the scalar field and
following reheating process, the amplitude of the high frequency modes
are determined by a combination of both the dynamics of inflation and the
reheating mechanism. Detailed numerical calculations of these processes
are therefore required in order to examine precisely the high frequency
gravitational waves which are the main subject of this paper.

In our calculation, we take into account the change of the effective
number of degrees of freedom and the neutrino anisotropic stress, both
of which cause a damping of the spectrum. It is notable that our
detailed calculations have also revealed a numerical error due to the
inaccurate treatment of the effect of neutrino anisotropic stress in
previous work (Ref. \cite{Watanabe}).  As a result, we obtain a wide
range spectrum which shows interesting features produced in the early
Universe, the inflationary phase and the reheating phase.  We should
therefore stress that this is the first work which calculates the
amplitude of the gravitational wave background accurately including all
of the factors that are definitely understood to affect it.  This makes
it possible to obtain an accurate spectral amplitude and is therefore
essential for accurately assessing the detectability of the
gravitational wave background.

The outline of our paper is as follows. In Sec. \ref{evoGW}, we begin
with the evolution equations expressing gravitational waves in terms of
metric tensor perturbations. Then we review the basics of the
gravitational wave background spectrum and evaluate the basic spectrum
shape analytically. In Sec.\ref{evoH}, we describe the physical
phenomena in the early Universe that are thought to have affected the
spectrum shape. In Sec.\ref{Ncal}, we present the results of our
numerical calculations of the gravitational wave background spectrum and
provide comparisons with theoretical predictions. We also show the
results at different reheating temperatures and different inflation
models. In Sec.\ref{Con}, we conclude with a discussion of the results
of Sec.\ref{Ncal}.

\section{The evolution of gravitational waves in the expanding Universe}
\label{evoGW}

Gravitational waves in the expanding Universe can be described as a
transverse-traceless part of the metric perturbation in a Friedmann
Robertson-Walker background. We consider the spatially flat case in
which the line element is given by
\begin{equation}
ds^2 =-dt^2+a^2(t)(\delta_{ij}+h_{ij})dx^idx^j,
\end{equation}
where the tensor perturbations $h_{ij}$ satisfy the
transverse-traceless conditions,
$h_{00}=h_{0i}=\partial^ih_{ij}=h^i_i=0$. Applying the metric of this
space-time to the perturbed Einstein equation $\delta G_{ij}=8\pi
G\delta T_{ij}$, we obtain the equation of motion for gravitational
waves
\begin{equation}
\ddot{h}_{ij}+3H\dot{h}_{ij}-\frac{1}{a^2}\nabla^2h_{ij}=16\pi
G\Pi_{ij},\label{heq1}
\end{equation}
where the over-dot describes the time derivative, $H$ is the Hubble
parameter, and $\Pi_{ij}$ is the transverse-traceless part of the
anisotropic stress of the energy-momentum tensor. The obvious
contribution from the anisotropic stress term is the damping effect due
to neutrino free-streaming, presented in
Refs. \cite{Boyle,Watanabe,Weinberg,Pritch}.  We take into account this
effect numerically solving the integro-differential equation shown in
Appendix \ref{appneu}. Although there may be other yet unknown
free-streaming particles which cause damping at higher frequencies, we
do not include such an uncertain component.

We now develop the basics of the gravitational wave background spectrum
assuming a perfect fluid $\Pi_{ij}=0$ for simplicity.  For convenience,
we decompose $h_{ij}$ into its Fourier modes with the comoving wave
number $\textbf{k}$ and denote the two independent polarization states
as
\begin{equation}
h_{ij}(t,\textbf{x})=\sum_{\lambda=+,\times}^{}\int\frac{d^3k}{(2\pi)^{3/2}}\epsilon_{ij}^{\lambda}
(\textbf{k})h_\textbf{k}^{\lambda}(t,\textbf{k})e^{i\textbf{k}\cdot\textbf{x}}.
\end{equation}
where the polarization tensors $\epsilon_{ij}^{+,\times}$ satisfies
symmetric and transverse-traceless condition and are normalized as
$\sum_{i,j}^{}\epsilon_{ij}^{\lambda}(\epsilon_{ij}^{\lambda^{\prime}})^*=2\delta^{\lambda\lambda^{\prime}}$.
Then Eq. (\ref{heq1}) becomes
\begin{equation}
\ddot{h}_\textbf{k}^{\lambda}+3H\dot{h}_\textbf{k}^{\lambda}+\frac{k^2}{a^2}h_\textbf{k}^{\lambda}=0.
\label{heq2}
\end{equation}
Note that, by neglecting the anisotropic stress term, the contribution
of other components in the Universe appears only through the second
term. This means we can follow the evolution of gravitational waves
with Eq. (\ref{heq2}) simply by computing the Hubble expansion rate,
the evolution of which is determined by the total energy densities of
the background components.

Let us briefly explain how the primordial gravitational waves evolve
in the expanding Universe. Initially, during inflation, the modes of
interest were deep inside the Hubble horizon ($k\gg aH$) and fluctuating
quantum-mechanically. During this phase, the amplitude of
$h_\textbf{k}^{\lambda}$ is determined according to Appendix
\ref{app1},
\begin{equation}
|h_\textbf{k}^{\lambda}|^2=\frac{16\pi G}{2ka^2}.
\label{initial}
\end{equation} 
As the Universe expands exponentially, the mode crosses outside the
horizon ($k<aH$). At this moment, we assume that the quantum
fluctuations become classical. Using the fact that the second term in
Eq. (\ref{heq2}) becomes dominant compared with the third term, one
can easily obtain the solution
\begin{equation}
h_\textbf{k}^{\lambda}\propto {\rm const.}
\label{sphorizon}
\end{equation}
After the end of inflation, the mode reenters the horizon ($k>aH$) and
starts to oscillate.  In this phase, the behavior of
$h_\textbf{k}^{\lambda}$ is described by the WKB solution (see Appendix
\ref{appWKB}),
\begin{equation}
h_\textbf{k}^{\lambda}\propto a^{-1}e^{\pm ik\tau},
\label{WKB}
\end{equation}
where $\tau$ is the conformal time defined by $d\tau\equiv dt/a(t)$.

We now define a dimensionless quantity to characterize the
strength of gravitational waves in terms of their energy density 
$\rho_{\rm GW}$ and the critical density of the
Universe $\rho_{c}\equiv3H^2/8\pi G$ \cite{Maggiore},
\begin{eqnarray}
\Omega_{\rm GW}&\equiv&\frac{1}{\rho_c}\frac{d\rho_{\rm GW}}{d\ln k}\nonumber\\
&=&\frac{1}{12}\left(\frac{k}{aH}\right)^2\frac{k^3}{\pi^2}\sum_{\lambda}^{}|h_\textbf{k}^{\lambda}|^2.
\label{OmegaGW}
\end{eqnarray}
In the second step, we have substituted the Fourier transformed form
of $\rho_{\rm GW}$, which is given by the 00-component of the stress-energy
tensor as follows,
\begin{eqnarray}
\rho_{\rm GW}&=&\frac{1}{64\pi Ga^2}\langle (\partial_{\tau}
 h_{ij})^2+(\vec{\nabla} h_{ij})^2 \rangle\nonumber\\
&=&\frac{1}{32\pi G}\int\frac{d^3k}{(2\pi)^3}\frac{k^2}{a^2}2\sum_{\lambda}^{}|h_\textbf{k}^{\lambda}|^2.
\label{rhoGW}
\end{eqnarray}
Here, we have assumed that the mode is inside the horizon and behaves
like Eq. (\ref{WKB}), which means that the conformal-time derivative of
the mode can be regarded as the same as the spatial derivative when taking
the spatial average, $\langle\cdots\rangle$.

One may define the tensor power spectrum ${{\cal P}}_T(k)$ in terms of the
primordial tensor power spectrum ${{\cal P}}_{T,{\rm prim}}(k)$ and
the tensor transfer function $T_T(k)$ as
\begin{equation}
{{\cal
    P}}_T(k)\equiv\frac{k^3}{\pi^2}\sum_{\lambda}^{}|h_\textbf{k}^{\lambda}|^2={{\cal
    P}}_{T,{\rm prim}}(k)T_T^2(k).
\end{equation}
Then the energy spectrum $\Omega_{\rm GW}$ can be rewritten as
\begin{equation}
\Omega_{\rm GW}=\frac{1}{12}\left(\frac{k}{aH}\right)^2{{\cal P}}_{T,{\rm prim}}(k)T_T^2(k).
\end{equation}
This expression is used in previous works because it is useful to
evaluate the spectrum by splitting the contributions to the spectrum
shapes into two parts, ${{\cal P}}_{T,{\rm prim}}(k)$ and $T_T(k)$.  The
spectrum shape is affected by the Hubble rate both when the mode leaves
the horizon during inflation and when the mode reenters the horizon
after inflation.  Since ${{\cal P}}_{T,{\rm prim}}(k)$ is the spectrum
at the end of inflation, it contains only the contribution from the
former, whereas $T_T(k)$ represents the time evolution of each mode
after the end of inflation, which means it contains only the
contribution from latter.  In most previous work, ${{\cal P}}_{T,{\rm
prim}}(k)$ has been evaluated analytically using the slow-roll
approximation.  We shall return to this in the next section.

We now show two simple examples of frequency dependence of the
spectrum -- a radiation-dominated and a matter-dominated case --
assuming the primordial spectrum is flat, ${{\cal P}}_{T,{\rm
    prim}}(k)\propto k^0$.  According to Eq. (\ref{sphorizon}),
$h_\textbf{k}^{\lambda}$ remains constant on superhorizon scales.
This means the transfer function can be written in terms of the
amplitude of $h_\textbf{k}^{\lambda}$ at horizon crossing,
$T_T^2(k)=|h_{\textbf{k},0}|^2/|h_{\textbf{k},\rm hc}|^2$, where the
subscript "0" denotes the present time and "hc" denotes the
horizon-crossing time. After the modes reenter the horizon,
$h_\textbf{k}^{\lambda}$ decreases inversely proportional to the scale
factor, so that $|h_{\textbf{k},0}^{\lambda}|=|h_{\textbf{k},\rm
  hc}^{\lambda}|(a_{\rm hc}/a_0)$, then we get $T_T^2(k)=a_{\rm
  hc}^2/a_0^2$. Therefore, the energy spectrum at the present time is
rewritten as $\Omega_{\rm GW,0}=(k^2/12a_0^2H_0^2){{\cal P}}_{T,{\rm
    prim}}(a_{\rm hc}^2/a_0^2)$. By using the relation $k=a_{\rm
  hc}H_{\rm hc}$, the value $a_{\rm hc}$ in $\Omega_{\rm GW,0}$ can be
converted into $k$ with the proportional relation between $a_{\rm hc}$
and $H_{\rm hc}$. During a radiation-dominated era, the Hubble rate
behaves as $H\propto a^{-2}$. This leads to $a_{\rm hc}\propto
k^{-1}$, then we obtain
\begin{equation}
\Omega_{\rm GW,0}\propto k^0.
\end{equation}
Similarly, during a matter-dominated era, $H\propto a^{-3/2}$ leads to
$a_{\rm hc}\propto k^{-2}$, then we obtain
\begin{equation}
\Omega_{\rm GW,0}\propto k^{-2}.
\end{equation}

\section{Physics of the early Universe and its effect on the spectrum}
\label{evoH}

As was seen in the previous section, the evolution of the Hubble
expansion rate is important to determine the shape of the
gravitational wave background spectrum. Here, we shall describe the
physics of the early Universe which affect the spectrum shape and
provide the equations necessary for calculating the evolution of the
Hubble rate.

\subsection{Slow-roll inflation}

In most models of inflation, the accelerated expansion is driven by a
scalar field $\phi$ whose equation of motion is given by
\begin{equation}
\ddot{\phi}+3H\dot{\phi}+V^{\prime}=0,
\end{equation}
where prime denotes derivative with respect to $\phi$. The Hubble
parameter is determined by the energy density of this scalar field,
$\rho_{\phi}=\dot{\phi}^2/2+V$, so that the Friedmann equation can be
written as
\begin{equation}
H^2=\frac{8\pi}{3m_{Pl}^2}\left(\frac{1}{2}{\dot{\phi}}^2+V(\phi)\right),
\end{equation}
where $m_{Pl}=1/\sqrt{G}$ is the Plank mass. For exponential expansion,
the energy density of the $\phi$ field should be dominated by its
potential energy, $\dot{\phi}^2/2\ll V$. This condition is often
characterized by the slow-roll parameters which are defined as
\begin{eqnarray}
&&\epsilon\equiv\frac{m_{Pl}^2}{16\pi}\left(\frac{V^{\prime}}{V}\right)^2,\\
&&\eta\equiv\frac{m_{Pl}^2}{8\pi}\frac{V^{\prime\prime}}{V}.
\end{eqnarray}
As long as the so-called slow-roll conditions, $\epsilon\ll 1$ and
$|\eta|\ll 1$, are satisfied, inflation continues keeping the Hubble
rate nearly constant and we can use the slow-roll approximation which
is a useful tool to theoretically study the fluctuations generated
during inflation.

In the slow-roll regime, the spectra of the primordial scalar
perturbation ${{\cal P}}_{S,{\rm prim}}(k)$ and tensor perturbation
${{\cal P}}_{T,{\rm prim}}(k)$ which are generated outside the horizon
can be derived as
\begin{eqnarray}
&&{{\cal P}}_{S,{\rm prim}}(k)=\frac{1}{\pi\epsilon}\left.\left(\frac{H}{m_{Pl}}\right)^2\right|_{k=aH},\\ 
&&{{\cal P}}_{T,{\rm prim}}(k)=\frac{16}{\pi}\left.\left(\frac{H}{m_{Pl}}\right)^2\right|_{k=aH},\label{PTprim}
\end{eqnarray}
where $H$ is evaluated at the time when the mode with wave number $k$
exits the horizon. To parametrize the deviation from scale invariance
of the tensor spectrum ${{\cal P}}_{T,{\rm prim}}(k)$, one may use the
tensor spectral index $n_T$ and its running $\alpha_T$ which can be
expressed in terms of the slow-roll parameters
\cite{Lidsey,Kosowsky},
\begin{eqnarray}
&&n_T(k)\equiv\frac{d\ln {{\cal P}}_{T,{\rm prim}}(k)}{d\ln k}\simeq-2\epsilon,\\
&&\alpha_T(k)\equiv\frac{dn_T}{d\ln k}\simeq 4\epsilon\eta-8\epsilon^2.
\end{eqnarray}
Then the power spectrum is represented in an expanded form in terms of
$\ln(k/k_0)$,
\begin{equation}
\ln\frac{{{\cal P}}_{T,{\rm prim}}(k)}{{{\cal P}}_{T,{\rm prim}}(k_0)}=n_T(k_0)\ln\frac{k}{k_0}+\frac{1}{2}\alpha_T(k_0)\ln^2\frac{k}{k_0}+\cdots,
\label{srspectrum}
\end{equation}
where $k_0$ is a pivot wavenumber. Also, the tensor-to-scalar ratio
can be written in terms of $\epsilon$,
\begin{equation}
r\equiv\frac{{{\cal P}}_{T,{\rm prim}}(k)}{{{\cal P}}_{S,{\rm prim}}(k)}\simeq16\epsilon.
\label{tsrasio}
\end{equation}
These expressions are used as a reference for comparison
with our numerical results in the subsequent section.

\subsection{Reheating process}

During reheating process that follows inflation, the energy of the
scalar field is transferred to radiation or other light particles
oscillating near the minimum of its potential.  Here, we consider the
perturbative decay of the scalar field \cite{reh1,reh2} and only
consider the interaction between the scalar field and light
fermions.  Then the equations for $\phi$ can be simply rewritten
including the decay term as \cite{Kolb,Kofman}
\begin{equation}
\ddot{\phi}+(3H+\Gamma)\dot{\phi}+V^{\prime}=0,\label{reheat1}
\end{equation}
where $\Gamma$ is the decay rate of $\phi$ particle, which depends on
oscillation frequency at the bottom of the potential.  Note that this
expression is valid only for the case where the mass of the interacting
field and the decay rate $\Gamma$ is sufficiently smaller than the
oscillation frequency of the $\phi$ field.  We do not deal with
nonlinear processes like preheating \cite{preh1,preh2}, which occurs if
the decay is bosonic and the parameters satisfy specific conditions
\cite{Kofman}.  Such processes may also generate gravitational waves at
high frequencies with a quite different spectrum shape, and may even
exceed the amplitude of the inflation-produced gravitational wave
background \cite{Easther,Juan,Dufaux}.  However, since we aim to provide
a conservative estimate of the spectrum, we only consider perturbative
processes which exist regardless of the chosen reheating scenario.

The additional set of equations to describe the reheating process
is as follows. The evolution equation for the radiation energy density
$\rho_r$ can be written as
\begin{equation}
\dot{\rho}_r+4H\rho_r=\Gamma\rho_{\phi}.\label{reheat2}
\end{equation}
The Friedmann equation is written in terms of the energy density of
the $\phi$ field and radiation field,
\begin{equation}
H^2=\frac{8\pi}{3m_{Pl}^2}(\rho_{\phi}+\rho_r).\label{reheat3}
\end{equation}
Therefore, using this sets of three equations,
Eqs. (\ref{reheat1}),(\ref{reheat2}) and (\ref{reheat3}), we can
simply calculate the Hubble rate following the evolution of both the
scalar field and the radiation field which are interacting with each
other.

Reheating ends around $\Gamma\sim H$ where the $\phi$ field begins to
decay exponentially. At this stage, the Universe becomes
radiation-dominated, so that we can define the reheating temperature
which can be related to $\Gamma$ \cite{Kolb},
\begin{equation}
T_{\rm RH}\sim
g_*^{-\frac{1}{4}}\left(\frac{45}{8\pi^3}\right)^{\frac{1}{4}}(m_{Pl}\Gamma)^{\frac{1}{2}},
\label{TRH}
\end{equation}
where $g_{*}(T)$ is the effective number of degrees of freedom.
This value is often used to characterize the reheating process.

\subsection{Thermodynamics of the radiation era}
\label{gstar}

After the Universe becomes radiation-dominated, the Hubble rate
basically behaves as $H\propto a^{-2}$. In the early stages of the
radiation-dominated era, particles in the Universe are relativistic
and contribute to the radiation energy density. However, when a
particle species becomes nonrelativistic as the Universe expands and
the temperature decreases, the contribution of the particle species to
the radiation drops exponentially and becomes negligible. This results
in the change of the radiation energy density and the evolution of the
Hubble rate, then induces a damping of the amplitude at high
frequencies. These processes for all particle species are treated in
the framework of thermodynamics as changes of the "effective number of
relativistic degrees of freedom," $g_*(T)$, which is introduced for
the convenience to evaluate the contribution of all relativistic
species as a whole.  We take into account this effect with a
formulation based on Ref. \cite{Kolb} which is briefly described
as follows.

During the radiation era, matter particles are in thermal equilibrium
and they obey Fermi-Dirac or Bose-Einstein statistics. In this case,
the energy density and the entropy density at temperature $T$ can be
expressed as
\begin{eqnarray}
&&\rho(T)=\frac{\pi^2}{30}g_*(T)T^4,\label{rhoT}\\
&&s(T)=\frac{2\pi^2}{45}g_{*s}(T)T^3,\label{sT}
\end{eqnarray}
Here $g_*(T)$ and $g_s*(T)$ can be written in the form of a summation
of the contributions of each species as
\begin{eqnarray}
g_*(T)=\sum^{}_ig_{*,i}(T)\left(\frac{T_i}{T}\right)^4,\\
g_{*s}(T)=\sum^{}_ig_{*s,i}(T)\left(\frac{T_i}{T}\right)^3,
\end{eqnarray}
where $g_{*,i}(T)$ and $g_{*s,i}(T)$ represent the contribution of the
$i$th particle species with mass $m_i$, degrees of freedom
$g_i$ and temperature $T_i$, and are given by
\begin{eqnarray}
g_{*,i}(T)=g_i\frac{15}{\pi^4}\int^{\infty}_{x_i}\frac{(u^2-x_i^2)^{\frac{1}{2}}}{e^u\pm
  1}u^2du,\\ g_{*s,i}(T)=g_i\frac{15}{\pi^4}\int^{\infty}_{x_i}\frac{(u^2-x_i^2)^{\frac{1}{2}}}{e^u\pm
  1}\left(u^2-\frac{x_i^2}{4}\right)du.
\end{eqnarray}
For this expression, we define $x_i\equiv m_i/T$ and $u\equiv E_i/T$,
where $E_i=\sqrt{|\textbf{p}_i|^2+m_i^2}$, and assume that the chemical
potential is negligible. The sign $+$ in the denominator denotes
fermions and $-$ denotes bosons.

With the help of entropy conservation, $sa^3={\rm const}$,
Eqs.(\ref{rhoT}) and (\ref{sT}) lead to $\rho\propto
g_*g_{*s}^{-4/3}a^{-4}$.  This means the behavior of the Hubble rate in
the radiation-dominated era is modified as $H\propto
g_*^{1/2}g_{*s}^{-2/3}a^{-2}$.  Let us briefly consider how the the
changes in $g_*$ and $g_{*s}$ affect the spectrum of the gravitational
wave background.  Referring back to the discussion at the end of Sec.
\ref{evoGW}, the energy spectrum is written as $\Omega_{\rm
GW,0}=(k^2/12a_0^2H_0^2){{\cal P}}_{T,{\rm prim}}(a_{\rm hc}^2/a_0^2)$.
Substituting the modified Hubble rate into the relation $k=a_{\rm
hc}H_{\rm hc}$, we obtain $a_{\rm hc}\propto g_{*,{\rm
hc}}^{1/2}g_{*s,{\rm hc}}^{-2/3}k^{-1}$.  This leads to $\Omega_{\rm
GW,0}\propto g_{*,{\rm hc}}^{-1/3}k^0$, where we assume $g_*=g_{*s}$,
which holds before the neutrino decoupling, for simplicity.  Therefore,
since $g_*$ is large at the early stage of the Universe when the
temperature is high enough for particles to be relativistic, the
spectrum is suppressed at the higher frequencies which enter the horizon
earlier.

Taking into account the improved evolution of the energy density, the
Friedmann equation for the Hubble rate can be expressed in terms of the
density parameters as
\begin{eqnarray}
H=H_0\left[\left(\frac{g_*}{g_{*0}}\right)\left(\frac{g_{*s}}{g_{*s0}}\right)^{-\frac{4}{3}}\Omega_r\left(\frac{a}{a_0}\right)^{-4}\right.\nonumber\\
\left.+\Omega_m\left(\frac{a}{a_0}\right)^{-3}+\Omega_{\Lambda}\right]^{\frac{1}{2}}\label{Hubble_g},
\end{eqnarray}
where $\Omega_r$, $\Omega_m$, and $\Omega_{\Lambda}$ are the density
parameters for radiation, matter and the cosmological constant,
respectively. Throughout this paper, we adopt
$\Omega_rh^2=4.31\times10^{-5}$ and the WMAP cosmological parameters
$h=0.732$, $\Omega_m=0.241$, and $\Omega_{\Lambda}=0.759$
\cite{Spergel}. The particles we assume are listed in Table I of
Ref. \cite{Watanabe}. Here, we consider only particles in the standard
model and do not include SUSY particles or any other exotic
particles. Including SUSY particles would double $g_*$ at above the TeV
scale and $\Omega_{\rm GW,0}(\propto g_{*,{\rm hc}}^{-1/3})$ would decrease by a
factor of $2^{-1/3}\simeq 0.8$ \cite{Chiba}.

\section{Spectrum of Primordial Gravitational Waves from Inflation}
\label{Ncal}

Before showing our numerical results, we briefly describe details of the
numerical calculation method. We first compute the background equations
for the Hubble rate using the procedure described in Sec. \ref{evoH}, and
then substitute it into the evolution equation of gravitational waves,
Eq. (\ref{heq2}), presented in Sec. \ref{evoGW}.  Note that this equation
is obtained by neglecting the anisotropic stress term, which can be set
to zero when neutrinos do not induce an anisotropic stress. We take into
account the contribution of the neutrino anisotropic stress by using
Eq. (\ref{heqneu}) after the time of neutrino decoupling.

For the calculation of the Hubble rate, we use
Eqs. (\ref{reheat1}),(\ref{reheat2}) and (\ref{reheat3}) during
inflation and reheating. Then we switch to Eq. (\ref{Hubble_g}) at the
point when the Universe becomes well dominated by radiation, $H\propto
a^{-2}$. The e-folding number of inflation, corresponding to the pivot
scale, is chosen in order to ensure smooth connection to the present
Hubble rate. We obtain its exact value by iteratively solving the
equations for inflation and reheating.

We start the calculation for each gravitational wave mode from a point
where the mode is well inside the horizon. The initial condition of the
mode is set to oscillate with the amplitude given in
Eq. (\ref{initial}). Then we solve the evolution of the gravitational
waves for each mode using Eq. (\ref{heq1}) and the evolution of the
Hubble rate. After the mode has reentered into the horizon and again
begun to oscillate rapidly, we replace its evolution with the WKB
solution of Eq. (\ref{WKB}) since it would be time consuming to treat the
oscillations numerically.  Note that we also use the WKB solution even
when the neutrino anisotropic stress is taken into account because the
effect is negligible if $k\gg aH$ (see Appendix \ref{appneu}).  Finally,
we compute the spectrum according to Eq. (\ref{OmegaGW}).

\subsection{Chaotic inflation}
\label{chao}

\begin{figure}
 \begin{center}
  \includegraphics[width=0.48\textwidth]{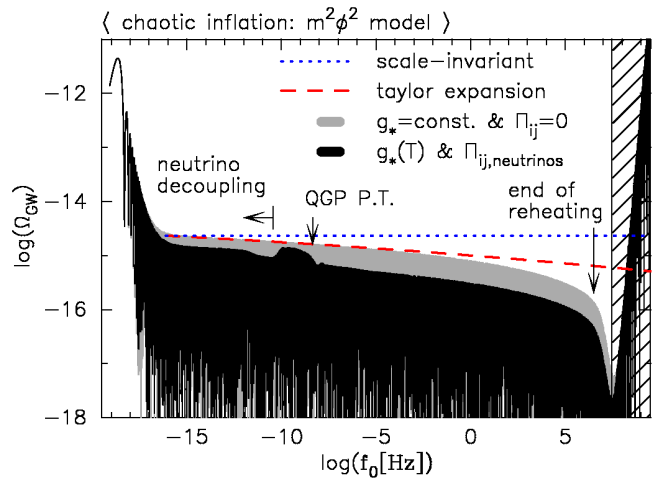} \caption{\label{phi2}
  The spectrum of the gravitational wave background generated during
  inflation with a quadratic potential $V=m\phi^2/2$, versus the
  gravitational frequency $f_0=k/2\pi$[Hz]. The black-line spectrum
  includes the effects of $g_*$ changes and the neutrino anisotropic
  stress, and the light gray one does not. Two lines are shown to
  compare our numerical result with theoretical prediction: the dotted
  curve represents a scale-invariant spectrum and the dashed line is the
  spectrum predicted under the slow-roll approximation, which is plotted
  according to Eq. (\ref{srspectrum}) with the spectral index $n_T(k_{\rm
  pivot})=-1.76\times 10^{-2}$ and its running $\alpha_T(k_{\rm
  pivot})=-3.11\times 10^{-4}$.  The shaded area corresponds to the
  spectrum due to quantum zero point fluctuations (see Sec. \ref{chao}).
  }
 \end{center}
\end{figure}

\begin{figure}
 \begin{center}
  \includegraphics[width=0.48\textwidth]{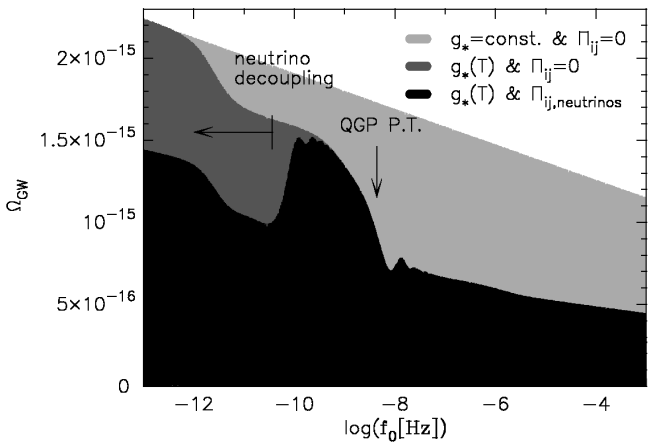} \caption{\label{zoom}
  Portion of the same spectrum of Fig. \ref{phi2}, focusing on the
  features induced by the change of $g_*$ and the neutrino anisotropic
  stress. Note that $\Omega_{\rm GW}$ is plotted on a linear scale. The
  black spectrum includes both contributions, the dark gray one includes
  only the effect of $g_*$ changes, and the light gray one does not
  include either.  }
 \end{center}
\end{figure}

First, we consider the case where the scalar field has a quadratic
potential,
\begin{equation}
V(\phi)=\frac{1}{2}m^2\phi^2.
\label{quad}
\end{equation}
We used the normalization of the scalar perturbations which is derived
by combining the constraints from recent observations \cite{Seljak},
\begin{equation}
{{\cal P}}_{S,{\rm prim}}(k_{\rm pivot}=0.05 {\rm Mpc}^{-1})=(2.45\pm 0.23)\times 10^{-9}.
\label{Ps}
\end{equation}
This fixes the value of the scalar field mass to be $m=1.64 \times
10^{13} $GeV. The decay rate is set to be $\Gamma=10^{-2}m$.

Figure \ref{phi2} shows the result of our numerical calculation. In
the lower frequency region, we see a well-known characteristics of the
spectrum as described in Sec.\ref{evoGW} -- that is the change of the
frequency dependence of the spectrum ($\Omega_{\rm GW}\propto k^{-2}$ to
$k^0$) at around $10^{-17}$Hz, which corresponds to the horizon
scale at matter-radiation equality. At frequencies above $10^{-17}$Hz,
the spectrum shows some interesting structural features, which we will
focus on for the remainder of this paper.

Let us first discuss the effect of inflation by paying attention to the
tilt of the spectrum. For comparison with previous works, we plotted the
Taylor-expanded spectrum derived in the slow-roll regime
(Eq. (\ref{srspectrum})). Note that, when comparing with the spectrum of
$g_*={\rm const}$ and $\Pi_{ij}=0$, our calculation is in good
agreement around the pivot wavenumber $k_{\rm pivot}=0.05 {\rm
Mpc}^{-1}$ which corresponds to $f_0\simeq 10^{-16.1}$Hz. On the other
hand, we see deviations from the Taylor-expanded spectrum at frequencies
above about $10^{-5}$Hz, which includes the frequency bands of LISA
\cite{lisa} and DECIGO/BBO \cite{decigo,bbo}.  This is because the
Taylor expansion around the pivot scale ($f_{\rm pivot}\simeq
10^{-16.1}$Hz) is no longer valid.  This can be avoided if one uses
Eq. (\ref{PTprim}) instead of Eq. (\ref{srspectrum}) as long as the
slow-roll condition is satisfied. However, at frequencies above
$10^6$Hz, where one finds the deviation becomes much larger, the
slow-roll condition is no longer satisfied. This causes a deviation from
the correct amplitude, and is unavoidable in the slow-roll
regime. Therefore, the slow-roll approximation overestimates the
amplitude of the spectrum at very high frequencies. Moreover, the use of
the Taylor-expanded form of the slow-roll approximation, which is used
widely in previous works, makes the deviation larger and also increases
the frequency range over which it occurs. Obviously, these deviations
are also present when the change of $g_*$ is taken into account.

Other characteristic features in the spectrum are produced after the end
of inflation.  When comparing with the spectrum which does not include
$g_*$ changes and the neutrino anisotropic stress, we find these two
effects cause damping over wide frequencies.  While the changes of $g_*$
produce damping of higher frequencies (above $10^{-12}$Hz) as explained
in Sec. \ref{gstar}, the neutrino anisotropic stress mainly causes
damping at lower frequencies ($10^{-17}$Hz to $10^{-10}$Hz).  At
frequencies where the two effects overlap, the spectrum shows features
produced by a combination of both the change of $g_*$ and the neutrino
anisotropic stress.  We present a close-up of the overlapped region in
Fig. \ref{zoom}.  Note that we additionally show the spectrum which only
includes the effect of $g_*$ changes to make it clear which effect
causes the damping at each frequency.  It may be seen that the stepwise
changes of $g_*$ induce a steplike shape in the spectrum and the
neutrino anisotropic stress only affects the modes which reenter the
horizon after the time of their decoupling at $2$MeV.  The magnitude of
the damping due to neutrino anisotropic stress may be derived
analytically to be 35.6$\%$ \cite{Weinberg}.  Note that our result does
not show the dip and peak around $f_0=0.8\times 10^{-10}$Hz
($k_0=5\times 10^{-10}$Hz) as seen in the results of
Ref. \cite{Watanabe}. We find that the feature does not arise for the
reason given in Appendix D of Ref. \cite{Watanabe}, but instead is the
result of inaccurate treatment of the source term. For details, see
Appendix \ref{appneu}.  \footnote{We thank Y. Watanabe (private
communication) for helping to clarify the reason for the difference
between our results and theirs.}

The change of the Hubble rate produces features in the
spectrum around the frequency of the mode that reenters the horizon
at the time of the change. One can relate the frequencies today to the
temperature of the Universe at the time when the modes reenter the
horizon as \cite{Kamionkowski}
\begin{eqnarray}
f_0&=&1.65\times 10^{-7}\times
\frac{1}{2\pi}\nonumber\\ &&\times\left[\frac{T_{\rm hc}}{1{\rm GeV}}\right]\left[\frac{g_{*s}(T_{\rm hc})}{100}\right]^{-\frac{1}{3}}\left[\frac{g_{*}(T_{\rm hc})}{100}\right]^{\frac{1}{2}} {\rm Hz}.
\label{freq}
\end{eqnarray}
For example, substituting the temperature of the quark-gluon phase
transition (called QGP P.T. in the figures), which is assumed to occur
at $180{\rm MeV}$, for $T_{\rm hc}$ yields $f_0\simeq
10^{-8}$Hz. Indeed, at around this frequency, we do find the large step
due to the sudden change of $g_*$ \cite{Watanabe}.  Also, the change of
the frequency dependence ($\Omega_{\rm GW}\propto k^0$ to $k^{-2}$) at
around $f_0\simeq 10^7$Hz corresponds to the reheating temperature
$T_{\rm RH}\simeq 2\times 10^{14} $GeV, which is derived by substituting
$\Gamma=10^{-2}m\simeq 10^{11} $GeV into Eq. (\ref{TRH}). It is notable
that the $k^{-2}$ dependence above $10^7$Hz is the reheating effect due
to the fact that the Hubble rate has the same evolution as a
matter-dominated Universe during reheating.

One thing we have to mention here is the significant rise at the highest
frequencies, which is proportional to $k^4$. These are the modes which
have not crossed outside the horizon during inflation, so that their
amplitude decrease as $\Omega_{\rm GW}\propto a^{-2}$ all the way
through keeping the gradient of the initial spectrum. Of course they may
not in truth exist because they does not "freeze in" as classical
fluctuations. One may regard them not as gravitational waves which have
been propagating since the epoch of inflation, but as quantum
perturbations of the space-time metric which are arising "just
now". \footnote{These modes are considered to be contributing to the
cosmological constant and tuned to be very small. However, for
reference, if they contribute to the total energy density of the
Universe, the energy density of the gravitational waves exceeds that of
the scalar field for $f_0\simg 10^{11}$ Hz, and the linear analysis
becomes invalid.}

\subsection{Low reheating temperature}

\begin{figure}
\begin{center}
\includegraphics[width=0.48\textwidth]{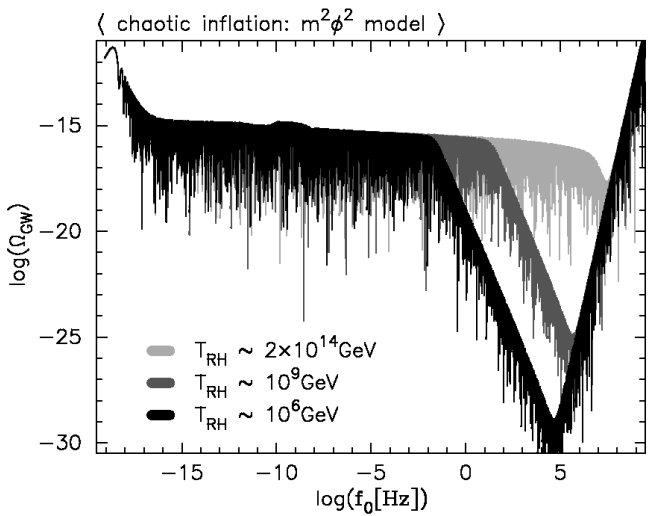}
\caption{\label{Gamma}The spectrum for different values of the
  reheating temperature for chaotic inflation. The lightest gray
  spectrum is the same as shown in Fig. \ref{phi2}, i.e. the decay
  rate is set to be $\Gamma=10^{-2}m$ which corresponds to $T_{\rm
    RH}\simeq 2\times 10^{14}$GeV. The dark gray one shows the case of
  $T_{\rm RH}\simeq 10^9 $GeV and the black one shows the case of
  $T_{\rm RH}\simeq 10^6 $GeV.}
\end{center}
\end{figure}

Here, we consider different reheating temperatures, referring to the
upper bound from the gravitino problem, $T_{\rm RH}<10^6-10^9$GeV
\cite{Kawasaki}. As we discussed in the previous subsection, the
reheating temperature determines the characteristic frequency of the
feature induced by reheating, so that different reheating temperatures
are expected to result in different spectrum shapes. Here, we consider
the case of $T_{\rm RH}=10^9 $GeV and $10^6 $GeV, which corresponds to
the decay rate $\Gamma\simeq 2\times10^{-11}m$ and $2\times10^{-17}m$.
The calculation is carried out with the same condition as that of
quadratic chaotic inflation except the value of the decay rate.

Shown in Fig. \ref{Gamma} is a comparison of the spectrum at different
reheating temperatures.  It can be seen that each spectrum has the
$k^{-2}$ dependence due to the reheating effect, although their length
are quite different. This is because a lower decay rate makes the
reheating time longer. Note that the edge at the frequency which
corresponds to the reheating temperature shifts to lower frequencies as
the reheating temperature decreases.  This behavior is consistent with
the estimation from Eq. (\ref{freq}) that the corresponding frequencies
for $T_{\rm RH}=10^6$GeV and $10^9$GeV are $10^{-2}$Hz and $10^{1}$Hz
which will be covered by DECIGO/BBO.  It is an exciting possibility to
determine the reheating temperature of the Universe by these experiments
\cite{Nakayama,Kuroyanagi}.

\subsection{Application to other inflation models}

\begin{figure}
  \begin{center}
    \includegraphics[width=0.48\textwidth]{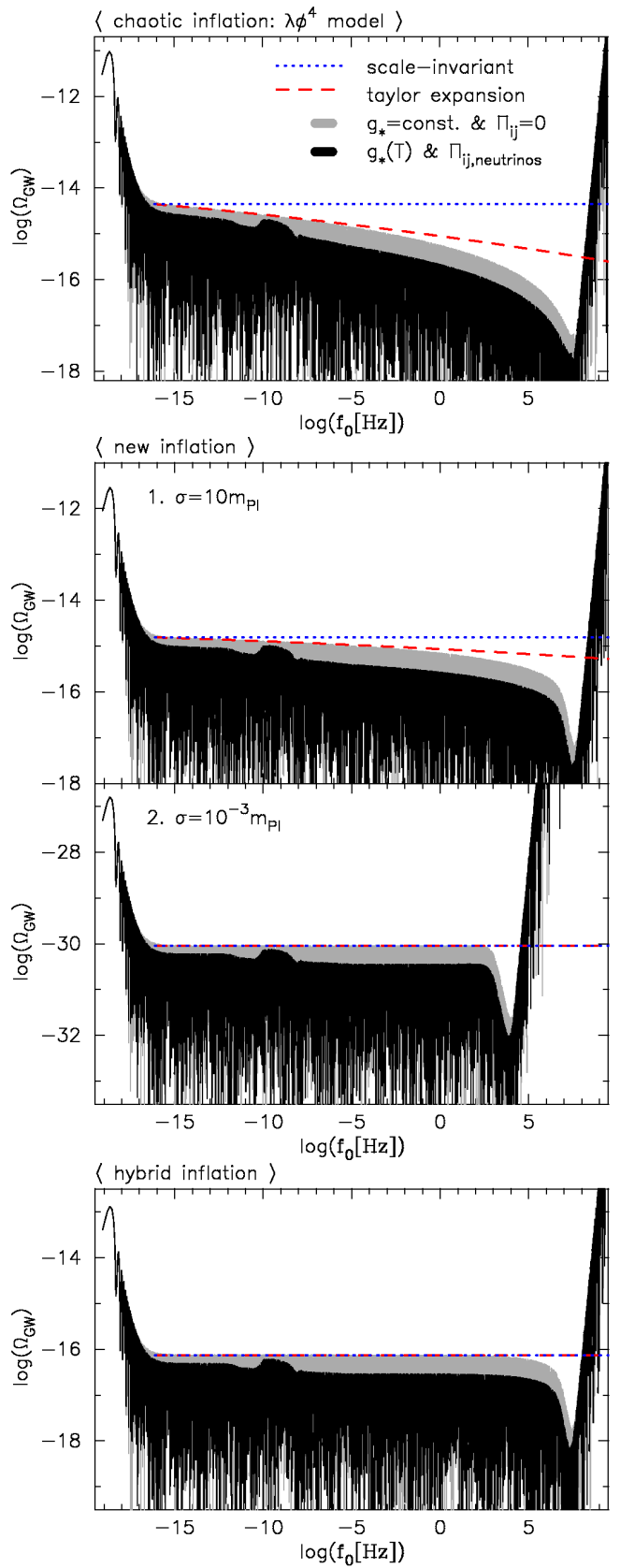}
    \caption{\label{models} The spectrum for different inflation potentials.}
  \end{center}
\end{figure}

We also repeat the calculations for other inflation models:
$\lambda\phi^4$, new inflation, and hybrid inflation. In each case, we
adopt the same normalization as in Eq. (\ref{Ps}) and set the decay
rate to be $\Gamma=10^{-2}\bar{m}$, where $\bar{m}$ is the effective
mass of the scalar field at the minimum of the potential. The results
for each model are summarized in Fig. \ref{models}. The values of
$n_T$, $\alpha_T$, which are required to draw a line of the slow-roll
prediction, and $r$ are listed in Table \ref{table}. Also, the
comparison between all models is shown in Fig. \ref{comp}.

\subsubsection{$\lambda\phi^4$ model}

Let us consider chaotic inflation with a quartic potential,
\begin{equation}
V(\phi)=\frac{1}{4}\lambda\phi^4,
\end{equation}
where the normalization gives $\lambda =1.78\times 10^{-13}$GeV. A
notable characteristic of this type of potential is that the decay
rate is not constant, because the oscillation frequency of $\phi$
depends on the amplitude of $\phi$,
i.e. $\bar{m}=c\sqrt{\lambda}\langle\phi\rangle$, where $c$ is a
constant of order 1 \cite{Morikawa,Shtanov}.

In the $\phi^4$ case, the Hubble rate behaves as in a
radiation-dominated Universe when the $\phi$ field oscillates about the
minimum of the potential.  Indeed, as one can find from the spectrum at
the top of Fig. \ref{models}, there is no matter-dominated phase
($k^{-2}$ dependence) soon after inflation. Therefore, for $\phi^4$
model, we cannot distinguish the stage of reheating and the subsequent
radiation-dominated era, so that the uncertainty of the decay rate does
not affect the resultant spectrum at all.  Note that the large gap at
high frequencies is only because of the large deviation from the
slow-roll approximation, which is also an interesting characteristic of
this potential. Again, as is seen from the figure, the Taylor-expanded
spectrum overestimates the spectrum at $f_0\simg 10^{-5}$ Hz.

\subsubsection{new inflation}

We now consider the original new inflation model which is 
based on the
Coleman-Weinberg potential \cite{Coleman},
\begin{equation}
V(\phi)=\frac{1}{4}\lambda\phi^4\left(\ln\frac{\phi}{\sigma}-\frac{1}{4}\right)+\frac{\lambda\sigma^4}{16}.
\end{equation}
The oscillation frequency is given by
$\bar{m}=\sigma\sqrt{\lambda}$. Here, we present the results for two
different parameter sets: (1) $\sigma=10m_{Pl}$, $\lambda=2.36\times
10^{-14}$GeV and (2) $\sigma=10^{-3}m_{Pl}$, $\lambda=7.27\times
10^{-12}$GeV. In case (1), we find a very similar spectrum to the case
of the $m^2\phi^2$ potential. By contrast, in case (2), the spectrum
is quite different. This difference arises from the difference in the
point in the potential at which the $\phi$ field rolls down. In case
(1), the scalar field rolled down where the potential is quadratic,
and this is why the spectrum has almost the same shape as that of the
$m^2\phi^2$ case.

In case (2), inflation occurs at a lower energy scale, and the decay
rate is set to be smaller, $\Gamma=10^{-6}\bar{m}$, in order to not
exceed the Hubble rate during inflation.  The remarkable features of
case (2) are the complete flatness of the spectrum and the smallness
of its amplitude. The flatness is due to the small gradient of the
potential because the initial value of $\phi$ is very near zero. The
smallness of the amplitude relates to the flatness of the spectrum,
about which we will explain later (see Sec. \ref{comparison}).

\subsubsection{hybrid inflation}

In the case of hybrid inflation, the potential is constructed with two
scalar fields $\phi$ and $\sigma$ \cite{Linde,Garcia},
\begin{equation}
V(\phi)=\frac{1}{4\lambda}(M^2-\lambda\sigma^2)^2+\frac{1}{2}m^2\phi^2+\frac{1}{2}g^2\phi^2\sigma^2.
\end{equation}
We consider the usual case where inflation is driven by the $\phi$
field and suddenly ends after the symmetry breaking of the $\sigma$
field (waterfall field). To satisfy this condition, we adopt the following parameters:
$\lambda=1$, $g=8\times 10^{-4}$, $m=1.5\times 10^{-7}m_{Pl}$,
$M=1.21\times 10^{16}$GeV. In this case, the oscillation at the
bottom of the potential mainly comes from the $\phi$ field, of which
the frequency is $\bar{m}_{\phi}=gM/\sqrt{\lambda}$, thus we can
assume that the radiation energy produced during reheating comes from
only the decay of the $\phi$ field.

The spectrum shown at the bottom of Fig. \ref{models} is nearly flat and
in good agreement with the Taylor-expanded spectrum. This is because the
quite gentle slope of the potential of the $\phi$ field results in an
almost constant value of the Hubble rate as long as the $\sigma$ field
triggers the sudden end of inflation. Thus, the spectrum has less
frequency dependence, which leads to less deviation from the Taylor
expansion. Also, interestingly, the value of $\alpha_T$ is positive,
which means $\eta>2\epsilon$. Both of these are the specific features of
hybrid inflation.

\subsubsection{Comparison}
\label{comparison}

\begin{table}
\begin{center}
\begin{tabular*}{0.48\textwidth}{@{\extracolsep{\fill}}lccc}
\hline
\hline
\rule[-3pt]{0pt}{14pt}Model & $n_T$ & $\alpha_T$ & $r$\\
\hline
\rule[0pt]{0pt}{12pt}$m^2\phi^2$ & $-1.76\times 10^{-2}$ & $-3.11\times 10^{-4}$ & 0.141\\
$\lambda\phi^4$ & $-3.40\times 10^{-2}$ & $-5.78\times 10^{-4}$ & 0.272\\
New (1) & $-1.16\times 10^{-2}$ & $-2.60\times 10^{-4}$ & $9.31\times 10^{-2}$\\
New (2) & $-6.83\times 10^{-18}$ & $-4.77\times 10^{-19}$ & $5.47\times 10^{-17}$\\
Hybrid & $-5.80\times 10^{-4}$ & $3.30\times 10^{-6}$ & $4.64\times 10^{-3}$\\[4pt]
\hline
\hline
\end{tabular*}
\caption{A summary of the values of the inflationary parameters $n_T$,
  $\alpha_T$, and $r$ for each model. These are evaluated at $k_{\rm
    pivot}=0.05 {\rm Mpc}^{-1}$.\label{table}}
\end{center}
\end{table}

\begin{figure}
\begin{center}
\includegraphics[width=0.48\textwidth]{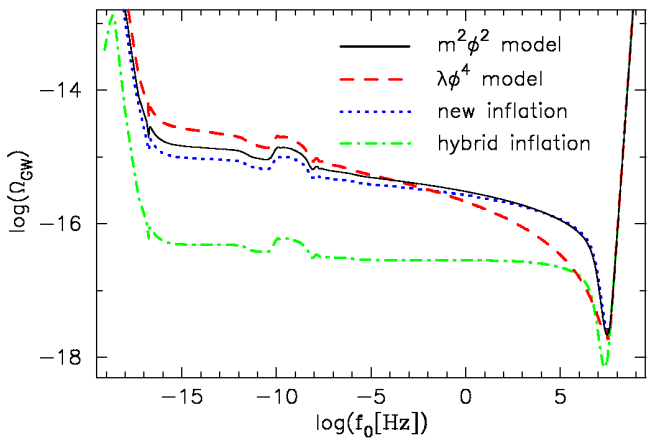}
\caption{\label{comp} The spectrum of gravitational wave background
  is shown for several inflation models. The solid curve shows the
  spectrum for the quadratic potential, the dashed curve for the quartic
  potential, the dotted curve for the new inflation potential with
  $\sigma=10m_{Pl}$, and the dot-dashed curve for hybrid inflation.}
\end{center}
\end{figure}

\begin{table*}
\begin{center}
\begin{tabular*}{0.98\textwidth}{@{\extracolsep{\fill}}lcccccc}
\hline
\hline
\rule[-3pt]{0pt}{15pt} & $10^{-16.1}$Hz & $10^{-5}$Hz & $10^{-3}$Hz &
 $10^{-1}$Hz & $10^2$Hz & ${\rm SNR}/\sqrt{T_{{\rm obs}}/10 {\rm yr}}$\\
\hline
\multicolumn{7}{c}{\rule[-5pt]{0pt}{15pt} Chaotic ($m^2\phi^2$) }\\
\hline
\rule[0pt]{0pt}{12pt}$g_*(T)$ \& $\Pi_{ij,{\rm neutrinos}}$ &
     $1.77\times 10^{-15}$ & $4.90\times 10^{-16}$ & $ 4.14\times
	     10^{-16} $ & $3.39\times 10^{-16}$ & $2.27\times 10^{-16}$
		     & 4.80\\
$g_*={\rm const}$ \& $\Pi_{ij}=0$ & $2.57\times 10^{-15}$ & $1.26\times
	 10^{-15}$ &$1.07\times 10^{-15}$ & $8.79\times 10^{-16}$ &
		     $5.92\times 10^{-16}$ & 12.0\\
Taylor-expanded slow-roll & $\cdots$ & $1.35\times 10^{-15}$ & $1.21\times
	     10^{-15}$ & $1.07\times 10^{-15}$ & $8.80\times 10^{-16}$ & 15.2
			 \\[4pt]
\hline

\multicolumn{7}{c}{\rule[-5pt]{0pt}{15pt} Chaotic ($m^2\phi^2$, $T_{\rm RH}=10^9{\rm GeV}$) }\\
\hline
\rule[0pt]{0pt}{12pt}$g_*(T)$ \& $\Pi_{ij,{\rm neutrinos}}$ &
     $2.75\times 10^{-15}$ & $4.90\times 10^{-16}$ &$ 4.01\times
	     10^{-16} $ & $3.17\times 10^{-16}$ & $9.12\times 10^{-18}$
		     & 4.53\\[4pt]
\hline

\multicolumn{7}{c}{\rule[-5pt]{0pt}{15pt} Chaotic ($m^2\phi^2$, $T_{\rm RH}=10^6{\rm GeV}$) }\\
\hline
\rule[0pt]{0pt}{12pt}$g_*(T)$ \& $\Pi_{ij,{\rm neutrinos}}$ &
     $2.88\times 10^{-15}$ & $4.90\times 10^{-16}$ &$ 3.94\times
	     10^{-16} $ & $1.51\times 10^{-17}$ & $7.60\times 10^{-24}$
		     & 0.20\\[4pt]
\hline

\multicolumn{7}{c}{\rule[-5pt]{0pt}{15pt} Chaotic ($\lambda\phi^4$) }\\
\hline
\rule[0pt]{0pt}{12pt}$g_*(T)$ \& $\Pi_{ij,{\rm neutrinos}}$ &
     $3.47\times 10^{-15}$ & $5.37\times 10^{-16}$ & $3.85\times
	     10^{-16}$ & $2.65\times 10^{-16}$ & $1.24\times 10^{-16}$
		     & 3.66\\
$g_*={\rm const}$ \& $\Pi_{ij}=0$ & $5.01\times 10^{-15}$ &$1.38\times
	 10^{-15}$ & $1.00\times 10^{-15}$ & $6.93\times 10^{-16}$ &
		     $3.27\times 10^{-16}$ & 9.46\\
Taylor-expanded slow-roll & $\cdots$ & $1.58\times 10^{-15}$ & $1.27\times
	     10^{-15}$ & $1.01\times 10^{-15}$ & $6.99\times 10^{-16}$ &
			 14.2\\[4pt]
\hline

\multicolumn{7}{c}{\rule[-5pt]{0pt}{15pt} New inflation ($\sigma=10m_{\rm pl}$)}\\
\hline
\rule[0pt]{0pt}{12pt}$g_*(T)$ \& $\Pi_{ij,{\rm neutrinos}}$ &
     $1.20\times 10^{-15}$ & $3.89\times 10^{-16}$ & $3.40\times
	     10^{-16}$ & $2.91\times 10^{-16}$ & $2.09\times 10^{-16}$
		     & 4.36\\
$g_*={\rm const}$ \& $\Pi_{ij}=0$ & $1.70\times 10^{-15}$ & $1.00\times
	 10^{-15}$ & $8.81\times 10^{-16}$ & $7.55\times 10^{-16}$ &
		     $5.47\times 10^{-16}$ & 11.0\\
Taylor-expanded slow-roll & $\cdots$ & $1.07\times 10^{-15}$ & $9.77\times
	     10^{-16}$ & $8.95\times 10^{-16}$ & $7.78\times 10^{-16}$ &
			 14.7\\[4pt]
\hline

\multicolumn{7}{c}{\rule[-5pt]{0pt}{15pt} New inflation ($\sigma=10^{-3}m_{\rm pl}$)}\\
\hline
\rule[0pt]{0pt}{12pt}$g_*(T)$ \& $\Pi_{ij,{\rm neutrinos}}$ &
     $7.08\times 10^{-31}$ & $3.55\times 10^{-31}$ &$3.54\times 10^{-31}$
	     & $3.54\times 10^{-31}$ & $3.51\times 10^{-31}$ & $4.95\times 10^{-15}$\\
$g_*={\rm const}$ \& $\Pi_{ij}=0$ & $1.00\times 10^{-30}$ & $9.02\times
	 10^{-31}$ & $9.01\times 10^{-31}$ & $9.01\times10^{-31}$ &
		     $8.97\times 10^{-31}$ & $1.27\times 10^{-14}$\\
Taylor-expanded slow-roll & $\cdots$ & $9.02\times 10^{-31}$ & $9.02\times 10^{-31}$ & $9.02\times10^{-31}$ & $9.02\times10^{-31}$ & $2.58\times 10^{-14}$\\[4pt]
\hline

\multicolumn{7}{c}{\rule[-5pt]{0pt}{15pt} Hybrid inflation}\\
\hline
\rule[0pt]{0pt}{12pt}$g_*(T)$ \& $\Pi_{ij,{\rm neutrinos}}$ & $
     5.75\times 10^{-17}$ & $ 2.88\times 10^{-17}$ & $ 2.85\times
	     10^{-17}$ & $2.84\times 10^{-17}$ & $2.77\times 10^{-17}$
		     & 0.41\\
$g_*={\rm const}$ \& $\Pi_{ij}=0$ & $ 8.32\times 10^{-17}$ &
	 $7.41\times 10^{-17}$ & $7.38\times 10^{-17}$ & $7.36\times
		 10^{-17}$ & $7.18\times 10^{-17}$ & 1.02\\
Taylor-expanded slow-roll & $\cdots$ & $7.41\times 10^{-17}$ & $7.38\times
	     10^{-17}$ & $7.36\times 10^{-17}$ & $7.34\times 10^{-17}$ &
			 1.05\\[4pt]
\hline
\hline
\end{tabular*}
\caption{ The density parameter of gravitational wave background
  $\Omega_{\rm GW}$ at $f_0=10^{-16.1},10^{-5}, 10^{-3},10^{-1},10^2$Hz for
  several inflation models, (1) calculated by fully numerically solving
  the evolution of gravitational waves including $g_*$ change and the
  neutrino isotropic stress, or (2) assuming $g_*={\rm const}$ and
  $\Pi_{ij}=0$, or (3) estimated by the Taylor-expanded form of the
  slow-roll approximation using Eq. (\ref{srspectrum}) with $g_*={\rm
  const}$ and $\Pi_{ij}=0$. In addition, the rightmost column shows the
  signal-to-noise ratios for DECIGO, which is aimed at
  $10^{-1}$Hz. \label{table2}}
\end{center}
\end{table*}

Finally, in Fig.\ref{comp}, we compare all models. In each case, the
effects of neutrino anisotropic stress and the $g_*$ changes appear of
course in the same way at the same position because these shapes are
formed well after the end of inflation. Needless to say, the differences
between the models arise due to the difference in how the scalar field
evolves in each potential during inflation.  The behavior of the Hubble
rate during reheating is also important as a model dependent
factor. Indeed, the reason for the unique shape of the $\lambda\phi^4$
model at high frequencies is because the Hubble rate evolves differently
than the other three models.

Note that the flatter the spectrum, the more the amplitude of
large-scale modes is suppressed. This behavior can be explained as
follows: From Eq. (\ref{tsrasio}), one can easily find ${\cal P}_{T,{\rm
prim}}=r{\cal P}_{S,{\rm prim}}$. Since we normalize the amplitude of
the spectrum using the value of ${\cal P}_{S,{\rm prim}}$, the smaller
scalar tensor ratio $r$ results in a smaller amplitude of the primordial
gravitational wave spectrum. In addition, $r$ can be rewritten as
$r\simeq -8n_T$, the tilt of the spectrum. This is equivalent to saying
that a flatter spectrum leads to a smaller amplitude primordial
gravitational wave spectrum. This can be confirmed by comparing the
values of $r$ or $n_T$ given in Table \ref{table} with the spectrum
amplitudes at the pivot scale (see the case of $g_*={\rm const}$ and
$\Pi_{ij}=0$), presented in the first column of Table \ref{table2}. The
extremely tiny value of $r$ in new inflation case (2) is the reason why
the amplitude of the spectrum in that case is quite small.

We also provide in Table \ref{table2} the detailed numerical values of
the spectrum at $f_0=10^{-5},10^{-3},10^{-1},10^2$Hz (including the
frequency bands of LISA \cite{lisa}, DECIGO/BBO \cite{decigo,bbo} and
LIGOII/LCGT \cite{ligo2,Kuroda}) for several inflation models.  For
comparison, the values of $\Omega_{\rm GW}$, calculated assuming $g_*={\rm
const}$ and $\Pi_{ij}=0$ or using the Taylor-expanded spectrum
predicted under the slow-roll approximation, Eq. (\ref{srspectrum}), are
also given. These clearly illustrate the suppression of the precise
values compared to those derived under these assumptions. For reference,
we calculate the predicted signal-to-noise ratio (SNR) using the noise
spectrum of DECIGO, which assumes a Fabry-Perot type space
interferometer and ten years of observing time ($T_{\rm obs}=10$yr)
\cite{Kudoh,Chiba}. This is shown in the rightmost column in Table
\ref{table2}. Note that we find large differences in SNR between results
from the spectrum we have obtained and the Taylor-expanded
spectrum. This clearly illustrates the reason why it is necessary to
evaluate the spectrum amplitude numerically. We find that the damping
effect due to the changes of $g_*$ decreases the SNR dramatically, and
also the assumption of the Taylor-expanded slow-roll overestimates the
SNR in some cases. The difference is more prominent for the models with
steeper potential such as chaotic inflation models. A detailed analysis
of the detectability will be done using the results of this paper in our
future work.

\section{Conclusion}
\label{Con}

Gravitational waves generated during inflation are expected to be a
powerful tool to probe the physics of the early Universe. In particular,
direct detection of gravitational waves would provide vital information
on their high frequency modes. In this paper, we have calculated the
spectrum of the gravitational wave background over a wide range of
frequencies (from $10^{-19}$ to $10^8$Hz) including the very high
frequencies where we expect to see a deviation from the slow-roll
approximation and the effect of reheating. This has become possible by
numerically solving the evolution of a scalar field from quantum states
during inflation until today taking into account the conversion of the
scalar field energy into radiation.  In addition, we have also taken
into account the contribution of neutrinos to the anisotropic stress and
the temperature dependence of the effective degrees of freedom
$g_*$. These two effects had already been investigated by
Ref. \cite{Watanabe}. However, we found that a correction is needed for
the treatment of the source term in their calculation.  Moreover their
calculation is limited in terms of frequency range, so that their result
is not relevant to the reheating process.  Therefore, this is the first
calculation that correctly includes all the factors which affect the
shape of the gravitational wave background spectrum over a wide range of
frequencies. This comprehensive treatment of the factors enables us to
make an accurate estimation of the amplitude of the spectrum, which
should be invaluable in making predictions for future experiments.

Our estimation shows that the Taylor-expanded spectrum of the slow-roll
prediction Eq.~(\ref{srspectrum}), which is analytically obtained and
used in many previous works, overestimates the amplitude of the spectrum
in some models. The deviation between our result and the analytical
prediction is especially evident at high frequencies ($f_0\simg 10^6$Hz)
because the slow-roll conditions are violated towards the end of
inflation. It is also notable that the deviation is found even in the
frequency region where the slow-roll conditions are satisfied (around
between $10^5$Hz and $10^{-6}$Hz in the $m^2\phi^2$ case). This is
because the Taylor expansion of the spectrum around the CMB scale is no
longer a good approximation at these frequencies.  However, since we do
not know the normalization of the scalar perturbations at much smaller
scales ($<{\rm kpc}$), it is impossible to carry out a Taylor expansion
about a high frequency point.  Therefore the deviation is inevitable as
long as we analytically evaluate the spectrum using the Taylor expansion
as in Eq.~(\ref{srspectrum}).  We strongly recommend numerical
calculation of spectra in order to discuss the detectability of the
gravitational wave background.

Moreover, we have shown that the spectrum has the characteristic
features produced after the end of inflation.  The effect of reheating
is observed in the spectrum as a dip at $\sim 10^8$Hz in the case where
the scalar field oscillates in a quadraticlike potential. This is
because the matter-dominated evolution of the Universe during reheating
causes the frequency dependence of $k^{-2}$. This feature is not seen in
the case of a quartic potential because of the difference in the
evolution of the Hubble rate.  In addition, the combination of the
effects due to changes in $g_*$ and the neutrino anisotropic stress is
observed between $10^{-17}-10^{-6}$Hz. It should be noted that changes
in $g_*$ decrease the amplitude by about half an order of magnitude at
higher frequencies, where the above-mentioned features of the early
Universe notably appear.  This damping effect becomes larger in case
where there exist SUSY particles or other exotic particles which cause a
change of $g_*$ \cite{Watanabe}. We have confirmed that this effect is
also important when evaluating the detectability.

Finally, we would like to conclude with some prospects for future
observations. When comparing the results of different inflation models,
we found that the amplitudes of the three models, both cases of chaotic
inflation and case (1) new inflation, are comparable around the
frequencies covered by future experiments like LISA, DECIGO/BBO and
LIGOII/LCGT. Thus, to distinguish these models, it may be necessary to
measure the tilt of the spectrum. Furthermore, interestingly, the
frequencies covered by future experiments cover the range in which the
spectrum shows an edge due to the effect of the low temperature
reheating. Hopefully, they will provide a new insight into understanding
of reheating.

Also, high frequency gravitational wave background around $10^8$Hz,
which the new design of detector suggested in Ref. \cite{Nishizawa} is
targeting, will be a probe of reheating process soon after inflation.
This is also the frequency at which we found the connecting point to
the $k^4$ dependence, beyond which the waves are considered to not
"freeze in" as classical fluctuations. However, the mechanism of
"freeze in" is yet to be uncovered. Therefore, through future direct
detection at around $10^8$Hz, we may obtain information not only on
reheating but also on classicalization of the quantum fluctuations.

\section*{Acknowledgments}

The authors are grateful to Takayuki Kaku for his contribution at the
early stage of this work, to Yuki Watanabe for useful correspondence,
and to Joanne Dawson for careful correction of the
manuscript. T.C. would like to thank Jun'chi Yokoyama for useful
discussion.  This research is supported by Grant-in-Aid for Nagoya
University Global COE Program, "Quest for Fundamental Principles in the
Universe: from Particles to the Solar System and the Cosmos", and
Grant-in-Aid for Scientific Research on Priority Areas No. 467 "Probing
the Dark Energy through an Extremely Wide and Deep Survey with Subaru
Telescope", from the Ministry of Education, Culture, Sports, Science and
Technology of Japan.  Also, this work was supported in part by
Grant-in-Aid for Scientific Research from JSPS [No.17204018(TC and NS)
and No.20540280(TC)] and from MEXT [No. 20040006(TC)] and in part by
Nihon University.

\appendix
\section{Quantization of tensor perturbations}
\label{app1}

To describe the initial condition of gravitational waves during
inflation, we need a quantum-mechanical treatment. Here, we do not go
into details but give a brief description of how to proceed (For
details, see Refs. \cite{Boyle,Lidsey}).

The quantization is carried out by introducing the rescaled variable
$v_\textbf{k}^{\lambda}(\tau,\textbf{k})\equiv
ah_\textbf{k}^{\lambda}/\sqrt{16\pi G}$ which satisfies the
commutation relations with its momentum
$\pi_\textbf{k}^{\lambda}(\tau,\textbf{k})=\partial
v_\textbf{k}^{\lambda}(\tau,\textbf{k})/\partial\tau$,
\begin{eqnarray}
&&[\hat{v}_{\textbf{k}}^{\lambda},\hat{\pi}_{\textbf{l}}^{\sigma\dagger}]=i\delta^{\lambda\sigma}\delta^{(3)}(\textbf{k}-\textbf{l}),\nonumber\\
&&[\hat{v}_{\textbf{k}}^{\lambda},\hat{v}_{\textbf{l}}^{\sigma}]=[\hat{\pi}_\textbf{k}^{\lambda\dagger},\hat{\pi}_{\textbf{l}}^{\sigma\dagger}]=0.
\label{comv}
\end{eqnarray}
We take the expression for the quantum operator
\begin{equation}
\hat{v}_\textbf{k}^{\lambda}(\tau,\textbf{k})=v_k(\tau,k)\hat{a}_\textbf{k}^{\lambda}+v_k^*(\tau,k)\hat{a}_{-\textbf{k}}^{\lambda\dagger},
\end{equation}
where the creation and annihilation operators,
$\hat{a}_\textbf{k}^{\lambda}$ and
$\hat{a}_{-\textbf{k}}^{\lambda\dagger}$, satisfy the commutation
relations
\begin{eqnarray}
&&[\hat{a}_{\textbf{k}}^{\lambda},\hat{a}_{\textbf{l}}^{\sigma\dagger}]=\delta^{\lambda\sigma}\delta^{(3)}(\textbf{k}-\textbf{l}),\nonumber\\
&&[\hat{a}_{\textbf{k}}^{\lambda},\hat{a}_{\textbf{l}}^{\sigma}]=[\hat{a}_\textbf{k}^{\lambda\dagger},\hat{a}_{\textbf{l}}^{\sigma\dagger}]=0.
\label{coma}
\end{eqnarray}
This follow from the Wronskian condition
\begin{equation}
v_k^*\frac{dv_k}{d\tau}-v_k\frac{dv_k^*}{d\tau}=-i.
\label{Wron}
\end{equation}
Then Eq. (\ref{heq1}) can be recast in terms of conformal-time
derivatives as
\begin{equation}
\frac{d^2v_k}{d\tau^2}+\left(k^2-\frac{1}{a}\frac{d^2a}{d\tau^2}\right)v_k=0.
\label{veq}
\end{equation}
Using the relation between the conformal time and the expansion rate
during inflation, $\tau\simeq -1/(aH)$ which is equivalent with
$da/d\tau\simeq-a/\tau$, the second derivative term can be rewritten as
$(d^2a/d\tau^2)/a\simeq 2/\tau^2$.  When considering the
short-wavelength limit $k/aH\rightarrow\infty$, we can neglect the term
of $1/\tau\simeq -aH$ which is much smaller than the $k$ term, and
obtain the asymptotic solution using the normalization of Eq. (\ref{Wron}),
\begin{equation}
v_k\rightarrow\frac{1}{\sqrt{2k}}e^{-ik\tau}.
\label{iniv}
\end{equation} 
This is the solution for the modes well inside the horizon during
inflation -- that is what we set as initial condition in our
calculation.

\section{The WKB solution}
\label{appWKB}
Here we present the derivation of the WKB solution, Eq. (\ref{WKB}).
First, we rewrite Eq. (\ref{heq2}) in terms of conformal time $\tau$,
\begin{equation}
\frac{d^2h_\textbf{k}^{\lambda}}{d\tau^2}+\left[\frac{2}{a}\frac{da}{d\tau}\right]\frac{dh_\textbf{k}^{\lambda}}{d\tau}+k^2h_\textbf{k}^{\lambda}=0.
\label{WKBtau}
\end{equation}
One may assume the form of the solution as
\begin{equation}
h_\textbf{k}^{\lambda}(\tau)=A(\tau)\exp[iB(\tau)],
\label{WKBsol}
\end{equation}
and consider the case in which the oscillation is very rapid compared to
the time variation of the amplitude, which means $dA/d\tau$ and
$d^2A/d\tau^2$ are much smaller than $dB/d\tau$.  By substituting
Eq. (\ref{WKBsol}) into Eq. (\ref{WKBtau}), we obtain two equations
arising respectively from the real part and the imaginary part,
\begin{eqnarray}
\frac{d^2A}{d\tau^2}-A\left(\frac{dB}{d\tau}\right)^2+\left[\frac{2}{a}\frac{da}{d\tau}\right]\frac{dA}{d\tau}+k^2A=0,\label{WKB1}\\
2\frac{dA}{d\tau}\frac{dB}{d\tau}+A\frac{d^2B}{d\tau^2}+\left[\frac{2}{a}\frac{da}{d\tau}\right]A\frac{dB}{d\tau}=0.\label{WKB2}
\end{eqnarray}
Using the assumption that $d^2A/d\tau^2\ll dB/d\tau$ and considering the
modes well inside the horizon $(da/d\tau)/a\ll k$, we obtain the solution
of $B(\tau)$ from Eq. (\ref{WKB1}),
\begin{equation}
B(\tau)=\pm k\tau+{\rm const.}
\end{equation}
The solution of $A(\tau)$ is derived by substituting this into Eq. (\ref{WKB2}),
\begin{equation}
A(\tau)\propto a^{-1}.
\end{equation}
Therefore we obtain the WKB solution,
\begin{equation}
h_\textbf{k}^{\lambda}(\tau)=\frac{C}{a}\exp(\pm ik\tau+\alpha).
\end{equation}
where $C$ and $\alpha$ are arbitrary constants.

\section{Anisotropic stress of free-streaming neutrinos}
\label{appneu}
\begin{figure}
 \begin{center}
  \includegraphics[width=0.48\textwidth]{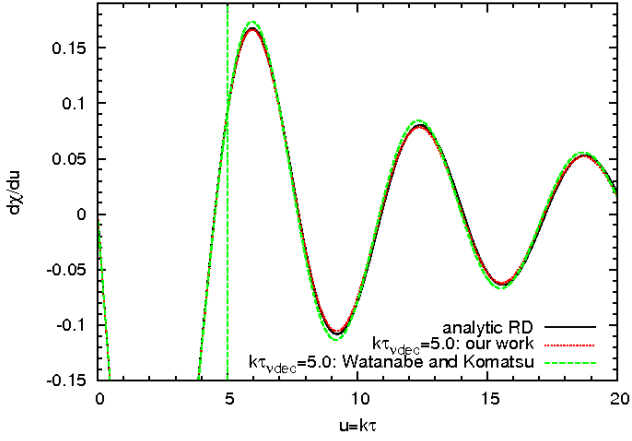}
  \caption{\label{neutrino1} Time evolution of $\chi^{\prime}(u\equiv
  k\tau)\equiv
  h_\textbf{k}^{\lambda\prime}(u)/h_\textbf{k}^{\lambda}(0)$ for the
  mode $k\tau_{\nu,{\rm dec}}= 5.0$, which entered the horizon before
  neutrino decoupling. The solid line is the analytical solution,
  $\chi^{\prime}=-j_1(u)$, and does not include the effect of neutrino
  anisotropic stress. The dotted line, which is almost indistinguishable
  from the analytical solution, represents our numerical result. The
  dashed line is an attempt to reproduce the numerical result of
  Ref. \cite{Watanabe}. The vertical dashed line represents the time of
  neutrino decoupling.}
 \end{center}
\end{figure}

\begin{figure}
 \begin{center}
  \includegraphics[width=0.48\textwidth]{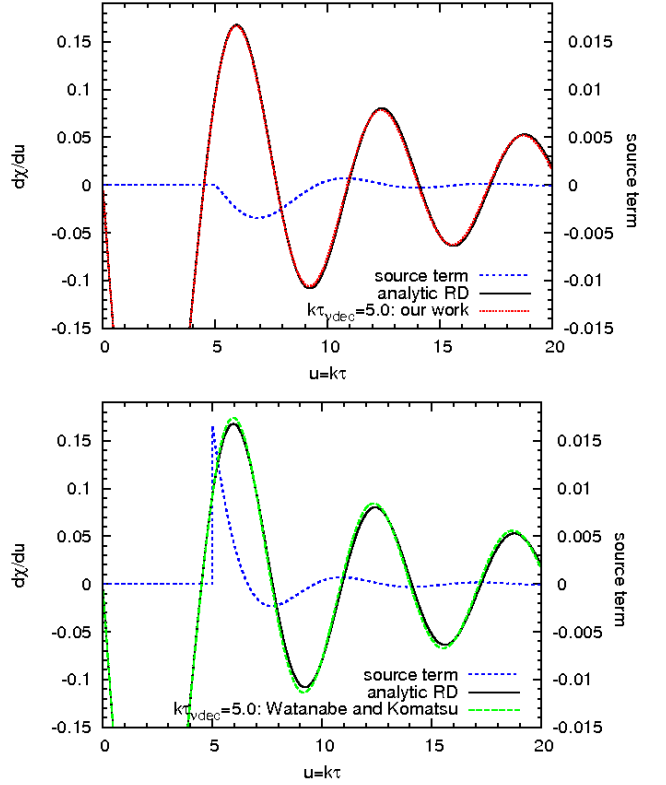}
  \caption{\label{neutrino2} Evolution of the source term, i.e. the
  right-hand side of Eq. (\ref{heqneu}). The upper figure shows our
  results and the lower shows the reproduced results of
  Ref. \cite{Watanabe}. The solid, dotted and dashed lines are the same
  as in Fig. \ref{neutrino1}, and the short-dashed line represents the
  source term. The artificially introduced discontinuity in the source
  term in the lower panel has been chosen by trial and error to recover
  the evolution of Ref. \cite{Watanabe}. Note that the scale of the
  vertical axis for the source term is 10 times larger than that for
  the others.}
 \end{center}
\end{figure}

Analytically solving the linearized Boltzmann equation for neutrinos, we
can calculate the spectrum including the effect of free-streaming
neutrinos by only evolving $h_\textbf{k}^{\lambda}$. According to
Ref. \cite{Weinberg} the evolution equation, Eq. (\ref{heq1}), can be
rewritten in the form of an integro-differential equation with using
conformal time $\tau$,
\begin{eqnarray}
\frac{d^2h_\textbf{k}^{\lambda}(\tau)}{d\tau^2}+\left[\frac{2}{a}\frac{da}{d\tau}\right]\frac{dh_\textbf{k}^{\lambda}(\tau)}{d\tau}+k^2h_\textbf{k}^{\lambda}(\tau)=-24f_{\nu}(\tau)\nonumber\\
\times\left[\frac{1}{a}\frac{da}{d\tau}\right]^2\int^\tau_{\tau_{\nu~{\rm
 dec}}}d\tau^\prime\left[\frac{j_2(k(\tau-\tau^{\prime}))}{k^2(\tau-\tau^{\prime})^2}\right]\frac{dh_\textbf{k}^{\lambda}(\tau^{\prime})}{d\tau},
\label{heqneu}
\end{eqnarray}
where $j_2$ is the spherical Bessel function of order 2. The function
$f_{\nu}(\tau)$ describes the fractional energy density of neutrinos,
which is defined using the total energy density $\rho$ and the energy
density of neutrinos $\rho_{\nu}$ as
\begin{equation}
f_{\nu}(\tau)\equiv\frac{\rho_{\nu}(\tau)}{\rho(\tau)}=\frac{f_{\nu,0}}{1+a(\tau)/a_{\rm eq}},
\end{equation}
where $f_{\nu,0}=\Omega_{\nu}/(\Omega_{\gamma}+\Omega_{\nu})=0.40523$.

Neutrinos do not induce an anisotropic stress at earlier epochs of the
Universe, because they are in equilibrium with electrons and photons
which are coupled to the nonrelativistic baryonic matter whose
anisotropic stress is negligible. After they decouple and begin
free-streaming, their anisotropic stress affects the evolution of the
gravitational waves as viscosity. As seen in Eq. (\ref{heqneu}), the
source term depends on the Hubble rate, which means the anisotropic
stress does not affect the mode when it is deep inside the horizon
($k\gg aH$). Thus, the damping effect does not arise at the
higher-frequency modes which were already inside the horizon at the time
of neutrino decoupling. Also, once the Universe becomes
matter-dominated, the contribution of neutrinos to the total energy
density of the Universe becomes sufficiently small ($f_{\nu}(\tau)\ll
1$), thus the damping effect does not appear in the modes which enter
the horizon during the matter-dominated era. Therefore, the damping
effect of the neutrino anisotropic stress is seen between the modes
which correspond to the time of neutrino decoupling ($f_0\sim
10^{-10}$Hz) and radiation-matter equality ($f_0\sim 10^{-17}$Hz).

Here, we explain the reason why our spectrum does not have the dip and
peak which are seen in the spectrum obtained by Watanabe and Komatsu
\cite{Watanabe}. In their paper, a small ripple is seen around the
frequency corresponding to the time of neutrino decoupling,
$f_0=0.8\times 10^{-10}$Hz ($k_0=5\times 10^{-10}$Hz), and is
interpreted as a result of assuming instantaneous decoupling. As seen in
Fig. \ref{zoom}, our result also shows a ripple at around the same
frequency, however it does not show a large dip and peak as observed in
their spectrum (see Fig. 5 of Ref. \cite{Watanabe}). The detailed reason
for the peak is explained in Ref. \cite{Watanabe} using a plot of the
time evolution of the amplified mode (see Fig. 10 of
Ref. \cite{Watanabe}). Their suggestion is that the mode gains energy
from neutrinos when the derivative of the mode, $\chi^{\prime}(u\equiv
k\tau)\equiv h_\textbf{k}^{\lambda\prime}(u)/h_\textbf{k}^{\lambda}(0)$,
is positive, so that it is amplified in the case where the mode enters
the horizon when $\chi^{\prime}>0$. However, we have reproduced the same
behavior by changing the source term, and found that the amplification
does not arise for the reason suggested in their paper.

The reproduced behavior of the derivative of the mode is plotted in
Fig. \ref{neutrino1}, which presents the same quantities as Fig. 10 of
Ref. \cite{Watanabe}. We also plot the result of our calculation in the
same figure. Note that the amplification which is seen in the reproduced
mode does not appear in our result. This difference is explained by the
behavior of the source term, which is presented in Fig. \ref{neutrino2}.
As seen in Fig. \ref{neutrino2}, the source term has an unnatural
discontinuous jump in the case where the mode is amplified. This
behavior can be reproduced when we calculate the source term by
replacing the lower limit of the integral in Eq. (\ref{heqneu}) with
zero and switch the source term to affect the evolution of gravitational
waves at the time of neutrino decoupling. Therefore the dip and peak
seen in Ref. \cite{Watanabe} are due to inaccuracy in computing the
source term. The validity of our calculation has been confirmed by the
author of Ref. \cite{Watanabe}, Y. Watanabe, via private communication.


\end{document}